\documentclass[sigconf,natbib=false]{acmart}

\usepackage{subcaption}
\usepackage{booktabs}
\usepackage{pgfplots}
\pgfplotsset{compat=1.18}

\AtBeginDocument{%
  }

\setcopyright{acmcopyright}
\copyrightyear{2023}
\acmYear{2023}
\acmConference[PASC'24]{The Platform for Advanced Scientific Computing (PASC) Conference}{June 3-5 2024}{Zürich, Switzerland}
\acmDOI{XXXXXXX.XXXXXXX}

\RequirePackage[
  datamodel=acmdatamodel,
  style=acmnumeric,
  ]{biblatex}

\newcommand{\Tor}{\mathbb{T}}

\newcommand{\cF}{{\mathcal F}}

\newcommand{\cO}{{\mathcal O}}

\newcommand{\bu}{{\bf u}}
\newcommand{\bk}{{\bf k}}
\newcommand{\bb}{{\bf b}}
\newcommand{\bB}{{\bf B}}

\newcommand{\bx}{{\bf x}}

\newtheorem*{maintheorem*}{Main Theorem}
\allowdisplaybreaks
\numberwithin{equation}{section}

\renewcommand{\i}{\ifmmode\mathit{\mathchar"7010 }\else\char"10 \fi}
\renewcommand{\j}{\ifmmode\mathit{\mathchar"7011 }\else\char"11 \fi}
\newcommand{\R}{\mathbb{R}}
\newcommand{\N}{\mathbb{N}}

\newcommand{\floor}[1]{\left\lfloor #1 \right\rfloor}

\copyrightyear{2024}
\acmYear{2024}
\setcopyright{acmlicensed}\acmConference[PASC '24]{Platform for Advanced Scientific Computing Conference}{June 3--5, 2024}{Zurich, Switzerland}
\acmBooktitle{Platform for Advanced Scientific Computing Conference (PASC '24), June 3--5, 2024, Zurich, Switzerland}
\acmDOI{10.1145/3659914.3659922}
\acmISBN{979-8-4007-0639-4/24/06}

\begin{document}

\title{Efficient Computation of Large-Scale Statistical Solutions to Incompressible Fluid Flows}


\author{Tobias Rohner}
\affiliation{%
  \institution{ETH Zürich}
  \streetaddress{Rämistrasse 101}
  \city{Zürich}
  \country{Switzerland}}
\email{tobias.rohner@sam.math.ethz.ch}

\author{Siddhartha Mishra}
\affiliation{%
  \institution{ETH Zürich}
  \streetaddress{Rämistrasse 101}
  \city{Zürich}
  \country{Switzerland}}
\affiliation{%
  \institution{ETH AI Center}
  \streetaddress{Andreasstrasse 5}
  \city{Zürich}
  \country{Switzerland}}
\email{siddhartha.mishra@sam.math.ethz.ch}

%\renewcommand{\shortauthors}{Tobias Rohner}

\begin{abstract}
  This work presents the development, performance analysis and subsequent optimization of a GPU-based spectral hyperviscosity solver for turbulent flows described by the three dimensional incompressible Navier-Stokes equations. The method solves for the fluid velocity fields directly in Fourier space, eliminating the need to solve a large-scale linear system of equations in order to find the pressure field. Special focus is put on the communication intensive transpose operation required by the fast Fourier transform when using distributed memory parallelism. After multiple iterations of benchmarking and improving the code, the simulation achieves close to optimal performance on the Piz Daint supercomputer cluster, even outperforming the Cray MPI implementation on Piz Daint in its communication routines. This optimal performance enables the computation of large-scale statistical solutions of incompressible fluid flows in three space dimensions.
\end{abstract}




%\received{XXX}
%\received[revised]{XXX}
%\received[accepted]{XXX}

\maketitle

\section{Introduction}
The flow of an incompressible fluid is described by the Navier-Stokes equations,
\begin{equation}
\label{eq:NS}
\begin{aligned}
\bu_t + \left(\bu \cdot \nabla \right) \bu + \nabla p &= \nu \Delta \bu, \quad (x,t) \in D \times [0,T], \\
{\rm div}~\bu &=0, \\
\bu(x,0) &= \bar{\bu}(x), \quad x \in D.
\end{aligned}
\end{equation}
Here, $\bu(x,t) \in \R^d$ is the velocity of the fluid, measured at the spatial location $x \in D \subset \R^d$ and time $t \in [0,T]$ and $p \in \R_+$ denotes the fluid pressure. The kinematic viscosity is denoted by $\nu$ and it scales inversely vis-à-vis the \emph{Reynolds number} $Re$ i.e., $\nu \sim \frac{1}{Re}$. It is well-known that for most fluids of interest \cite{Frisch1995}, particularly in the atmosphere and the ocean as well as in flows of practical engineering interest, the Reynolds number can be very high. Hence, one is interested in the regime $\nu \rightarrow 0$, which also corresponds to the zero-viscosity limit of the Navier-Stokes equations. 

Formally, when we assume periodic boundary conditions by setting $D = \Tor^d$ to be the $d$-dimensional periodic Torus, letting $\nu \rightarrow 0$ in \eqref{eq:NS}, one obtains the well-known Euler equations for an ideal, incompressible fluid,
\begin{equation}
\label{eq:euler}
\begin{aligned}
\bu_t + \left(\bu \cdot \nabla \right) \bu + \nabla p &= 0, \quad (x,t) \in D \times [0,T], \\
{\rm div}~\bu &=0, \\
\bu(x,0) &= \bar{\bu}(x), \quad x \in D.
\end{aligned}
\end{equation}

\subsection{The Role of Turbulence}
Turbulence \cite{Frisch1995} is loosely defined as the presence of energetic eddies that span a very large range of spatial and temporal scales, even if the initial data is only varying at a single scale. This spontaneous appearance of multiple scales is a manifestation of the nonlinearities in the momentum equation in \eqref{eq:NS} and implies a cascade of input energy into smaller and smaller scales. 

Turbulence is the principal obstacle in the availability of global well-posedness results for the Euler and Navier-Stokes equations. It is also responsible for the possible lack of convergence of numerical methods as a large number of scales, corresponding to wave numbers of $\approx Re^{\frac{3}{4}}$ need to be resolved, making the computational cost of a direct numerical simulation (DNS) of the Navier-Stokes equations prohibitive. In particular, these facts automatically imply that one can only expect convergence with respect to grid size for the Navier-Stokes equations when grid sizes smaller than $\ll\nu^{\frac{3}{4}}$ are considered. For the Euler equations, as $\nu =0$, this also implies that classical numerical methods will not converge to a weak solution. This fact is already computationally verified in two spatial dimensions with rough initial data \cite{LM1,LMP1}. Thus, it is unclear what exactly a numerical method actually computes when approximating the Euler and Navier-Stokes equations (for high Reynolds numbers) at a given resolution. 

\subsection{Statistical Solutions}
The above discussion clearly brings out the fact that the current solution concepts of weak and strong solutions for the Navier-Stokes (and Euler) equations are inadequate. They may not be globally well-posed and numerical methods may not converge to them on mesh refinement. This also provides the rationale for the search of alternative, more suitable, solution paradigms for incompressible flow. 

One such alternative is the concept of measure-valued solutions, first introduced in \cite{DM1}. Herein, the solutions are sought for as \emph{Young measures} or space-time parametrized probability measures. One can think of them as assigning a probability distribution (pdf) at each point in space-time. Measure-valued solutions exist globally \cite{DM1} and can be realized as limits of popular numerical methods such as the spectral viscosity method \cite{LM1}. However, they are not unique. This is due to the fact that there is no information on correlations across different points in space.

Adding information about all possible multi-point correlations in an attempt to recover uniqueness leads to the paradigm of \emph{Statistical Solutions} \cite{FTbook,FLM1} and references therein. Statistical solutions are time-parametrized probability measures on the underlying function space of square-integrable velocity fields. Their time evolution can be written as an infinite system of nonlinear differential equations, each evolving a particular moment in terms of other moments. 

Statistical solutions provide a language to express turbulence mathematically in terms of solutions of Euler and Navier-Stokes equations \cite{FTbook}. Moreover, they also provide a natural framework for quantifying the uncertainties that are inherent to fluid flows \cite{FLM1}. Recently, the equivalence of two different definitions of statistical solutions was shown in \cite{FMW1}. Moreover, it was also shown that under the assumptions analogous to (weaker than) those used by Kolmogorov to derive his famous K41 phenomenological theory of turbulence, one can prove that statistical solutions of the Navier-Stokes equations \eqref{eq:NS} converge to statistical solutions of the Euler equations \eqref{eq:euler} as $\nu \rightarrow 0$.

\section{Simulation}
The simulation developed in the scope of this work is written in C++ using CUDA to realize optimized compute kernels to run on GPGPUs. For performance considerations, the computational domain is restricted to the $d$-dimensional torus $\Tor^d$ discretized by a uniform rectangular grid with mesh width $\Delta = \frac{1}{N}$. Large simulations are distributed over multiple compute nodes along a single dimension of the grid, while communication of data between nodes is implemented using MPI.

\subsection{Computing Statistical Solutions}
Given that statistical solutions are probability measures on the infinite dimensional space of square-integrable velocity fields, the challenge of computing statistical solutions is formidable. However, in recent papers \cite{FLMW1,Lye2020,LMP1}, a Monte Carlo ensemble averaging algorithm to compute statistical solutions for both compressible as well as incompressible flows was proposed. Let the initial uncertainty in \eqref{eq:euler} (for definiteness) be modeled in terms of a probability measure $\bar{\mu} \in {\rm Prob}((L^2(\Tor^d)^d))$. The aim is to find a suitable approximation to the statistical solution $\mu_t$, whose correlation marginals satisfy the corresponding moment equations. To approximate $\mu_t$, \cite{FLMW1} and \cite{LMP1} proposed the following algorithm: \\\\
\noindent {\bf Algorithm $1$:}
\begin{itemize}
\item [1.]  Given initial measure $\bar{\mu}$, find $M$ Monte Carlo samples $\bar{\bu}_i(\omega)$ such that $\bar{\mu} \approx \frac{1}{M}\sum_{i=1}^M \delta_{\bar{\bu}_i(\omega)}$.
\item [2.]  $\forall \omega$, evolve $\bar{\bu}_i(\omega)$ with suitable numerical method, at mesh resolution $\Delta$, to obtain $\bu^{\Delta}_{i}(t)$.
\item [3.]  Define approximate statistical solution by the \emph{empirical measure}:
$\mu^{\Delta,M}_t = \frac{1}{M} \sum_{i=1}^M \delta_{\bu^{\Delta}_i(t)}$.
\end{itemize} \null
Note the possibility of two independent parallelization strategies for accelerating the simulation. One approach computes multiple samples $\bu_i^{\Delta}(t)$ in parallel, while the other approach parallelizes the evaluation of a single sample. Our simulation supports both strategies, as well as using them in combination. Due to the embarrassingly parallel nature of Monte Carlo sampling, we will only discuss parallelization of the evaluation of a single sample here.

\subsection{Spectral Hyper-Viscosity Method}
Note that in Algorithm $1$ the number of Monte Carlo samples $M$ should ideally scale with $M \propto N^2 = \frac{1}{\Delta^2}$. It is therefore of utmost importance that the computation of a single sample is implemented as efficiently as possible. Fourier spectral methods are an obvious choice of numerical method given our toroidal domain. In Fourier space, enforcing the divergence-free constraint of \eqref{eq:euler} reduces to a projection of each Fourier mode onto divergence-free vector fields. This circumvents the expensive computation of the pressure field reducing the computational complexity of the solver from $\cO(N^{2d+1})$ to $\cO(N^{d+1}\log N)$ where $d$ is the spatial dimension.

We consider the following spatial discretization of the incompressible Navier-Stokes equations \cite{LMP1}
\begin{equation} \label{eq:shv_disc}
    \begin{cases}
        \partial_t \bu^{\Delta} + \mathcal{P}_N(\bu^{\Delta}\cdot\nabla \bu^{\Delta}) + \nabla p^{\Delta} &= \varepsilon_N|\nabla|^{2s}(Q_N*\bu^{\Delta}) \\
        \nabla\cdot \bu^{\Delta} &= 0 \\
        \bu^{\Delta}|_{t=0} &= \mathcal{P}_N\bu_0
    \end{cases}
\end{equation}
where $\mathcal{P}_N$ is the spatial Fourier projection operator mapping a function $f(x,t)$ to its first $N$ Fourier modes: $\mathcal{P}_N = \sum_{|k|_{\infty}\leq N} \hat{f}_k(t)e^{ik\cdot x}$. We additionally introduce the hyperviscosity parameter $s \geq 1$. When simulating the vanishing viscosity limit of the Navier-Stokes equations, this parameter can be used to fine-tune the dampening of the Fourier modes allowing for a large part of the spectrum to be free of dissipation.
The viscosity term we use for the stabilization of the solver consists of a possibly resolution dependent viscosity $\varepsilon_N$ and a Fourier multiplier $Q_N$ controlling the strength at which different Fourier modes are dampened. This allows us to remove the dampening for the low frequencies, while applying some diffusion to the problematic higher ones. The Fourier multiplier $Q_N$ is of the form
\begin{equation}
    Q_N(\mathbf{x}) = \sum_{\mathbf{k}\in\mathbb{Z}^d, |\mathbf{k}|\leq N} \hat{Q}_{\mathbf{k}} e^{i\mathbf{k}\cdot\mathbf{x}}.
\end{equation}
In order to have convergence, the Fourier coefficients of $Q_N$ need to fulfill \cite{Tadmor1989,Tadmor2004,LMP1}
\begin{equation}
    \hat{Q}_\bk = 0 \mbox{ for } |\bk|\leq m_N, 1-\left(\frac{m_N}{|\bk|}\right)^{\frac{2s-1}{\theta}} \leq \hat{Q}_\bk \leq 1
\end{equation}
where we have introduced an additional parameter $\theta > 0$. The quantities $m_N$ and $\varepsilon_N$ are required to scale as
\begin{equation}
    m_N \sim N^{\theta}, \varepsilon_N \sim \frac{1}{N^{2s-1}}, 0 < \theta < \frac{2s-1}{2s}.
\end{equation}

The authors of \cite{LM2} show convergence of the above numerical method for a large class of initial conditions in the case of the two-dimensional incompressible Euler equations. For three dimensions no such result is available, but experimental evidence performed by the authors suggest that the results persist for three dimensions as well.

\subsection{Implementation}

Applying the Fourier transform to \eqref{eq:shv_disc} and using the divergence-free constraint to replace the pressure by its exact solution yields
\begin{equation}
    \partial_t\hat{\bu}_\bk^{\Delta} = \left( 1 - \frac{\bk\bk^T}{|\bk|^2} \right) \cdot \hat{\bb}_\bk
\end{equation}
where $\hat{\bb}_\bk = -i\bk^T \cdot \cF\left[\bu^{\Delta} \otimes \bu^{\Delta}\right]_\bk$. Note that the equation enforces the divergence-free constraint by orthogonally projecting the $\hat{\bb}_\bk$ onto vector fields fulfilling $i\bk^T \cdot \hat{\bb}_\bk = 0$. This represents a pointwise operation in Fourier space resulting in the superior computational complexity of this spectral method over other classical methods.

The nonlinear term $\cF\left[\bu^{\Delta} \otimes \bu^{\Delta}\right]$ is computed in physical space and uses the well known 2/3-dealiasing rule \cite{SO1} to ensure stability of the solution. The algorithm to compute the time derivative $\partial_t \hat{\bu}_\bk$ can thus be described by the following steps, each represented by a compute kernel in the code itself: \\\\
\noindent {\bf Algorithm $2$:}
\begin{itemize}
    \item[1.] Pad $\hat{\bu}^{\Delta}$ with zeroes to size $\frac{3}{2}N$
    \item[2.] Compute $\bu^{\Delta} = \cF^{-1}\left[\hat{\bu}^{\Delta}\right]$
    \item[3.] Compute $\bB = \bu^{\Delta} \otimes \bu^{\Delta}$
    \item[4.] Compute $\hat{\bB} = \cF\left[\bB\right]$
    \item[5.] Dealias by removing the upper third of frequencies in $\hat{\bB}$
    \item[6.] Use $\hat{\bB}$ to compute $\partial_t\hat{\bu}_\bk^{\Delta}$
\end{itemize} \null
Note that all kernels in Algorithm $2$ have very low and (almost) constant operational intensity in the mesh resolution $N$. This suggests that reducing the number of memory accesses is crucial for obtaining a highly performant code. We do this with a combination of algorithmic improvements, kernel fusion, and blocking as detailed later in this paper.

Given the $\cO(N^d)$ memory requirements to capture the current state of the simulated flow field $\bu^{\Delta}$, the need to distribute computations across multiple compute nodes arises quite early when striving for higher resolution simulations. (A single GPU on Piz Daint is able to contain a simulation of size $256^3$). Note that the only nonlocal kernels present in Algorithm $2$ are the forward and backward Fourier transforms, implying that the memory layout should be optimized towards their most efficient computation. Computing FFTs over distributed data, if desired to be efficient, is a nontrivial task \cite{Duy2014}. We desire to minimize the amount of inter-node communication required by the algorithm, as we expect all the compute kernels to be memory bound. Hence, the optimal split of the data over compute nodes is given by a \emph{slab decomposition} requiring only two transpose operations per FFT. Given all kernels, with minimal modifications to the code, can work with transposed data, the algorithm only needs a single transpose operation per FFT.

% \begin{figure}
%     \begin{tikzpicture}
%         \fill[rounded corners=15, fill=black!30!green!25] (0,0) rectangle (8,5);
%         \node[anchor=north] at (4,5) {\verb|MPI_COMM_WORLD|};
%         \fill[rounded corners=15, fill=black!25] (0.25,0.25) rectangle (3.25,4.25);
%         \node[anchor=north] at (1.75,4.25) {\verb|SUB_COMM_1|};
%         \fill[rounded corners=15, fill=black!25] (4.75,0.25) rectangle (7.75,4.25);
%         \node[anchor=north] at (6.25,4.25) {\verb|SUB_COMM_M|};
%         \node at (4.05,2.25) {\scalebox{2}{$\cdots$}};
%         \fill[rounded corners=15, fill=black!30!purple!25] (0.5,3.5) rectangle (3,2.3);
%         \node at (1.75,2.9) {Node 1};
%         \fill[rounded corners=15, fill=black!30!purple!25] (0.5,1.7) rectangle (3,0.5);
%         \node at (1.75,1.1) {Node N};
%         \node at (1.75,2.1) {$\vdots$};
%         \fill[rounded corners=15, fill=black!30!purple!25] (5,3.5) rectangle (7.5,2.3);
%         \node at (6.25,2.9) {Node 1};
%         \fill[rounded corners=15, fill=black!30!purple!25] (5,1.7) rectangle (7.5,0.5);
%         \node at (6.25,1.1) {Node N};
%         \node at (6.25,2.1) {$\vdots$};
%     \end{tikzpicture}
%     \caption{Typical communication structure of a simulation. Monte Carlo samples are distributed over $M$ independent subcommunicators, each computing a single sample at the time using $N$ compute nodes.}
%     \label{fig:mpi_setup}
% \end{figure}
%The typical communication structure used by a simulation is shown in Figure \ref{fig:mpi_setup}.
The typical MPI setup of a simulation looks as follows: The Monte Carlo samples are distributed over $M$ independent subcommunicators, each computing a single sample at a time parallelized over $N$ compute nodes. This strategy enables optimal utilization of compute resources. By taking $N$ to be the minimal amount of compute nodes needed to fit a simulation of a single sample we maximize the degree of parallelization over Monte Carlo samples. As each sample can be computed independently, this provides us with a near optimal scaling of our simulation on massively parallel hardware.

Although our simulator supports arbitrary time-stepping schemes, we generally use the third-order strong stability preserving Runge-Kutta described by
\begin{equation}
    \begin{aligned}
        u^{(1)} &= u(t) + \Delta t \partial_t u(t) \\
        u^{(2)} &= \frac{3}{4}u(t) + \frac{1}{4}u^{(1)} + \frac{1}{4}\Delta t \partial_tu^{(1)} \\
        u(t + \Delta t) &= \frac{1}{3}u(t) + \frac{2}{3}u^{(2)} + \frac{2}{3}\Delta t \partial_tu^{(2)}.
    \end{aligned}
\end{equation}
Contrary to other SSP Runge-Kutta schemes, the third-order one is energy-diminishing in the context of linear hyperbolic systems\cite{Tadmor98}. This means that as long as the CFL condition is respected, the simulation is much more likely to stay stable even with nonlinear PDEs such as the incompressible Euler and Navier-Stokes equations.

\subsection{In-Situ Processing}

Our simulator's flexible IO system allows us to perform most post-processing of the data in-situ saving a massive amount of IO bandwidth and disk space. Flexibility is achieved by abstracting away the process of storing data to disk by introducing the concept of a \verb|Writer|. Each writer provides the simulation with a list of time points at which it processes the current simulation state and stores the result to disk. Furthermore, a simulation can contain an arbitrary number of different writers each one storing data at different time intervals. Currently supported operations by writers include:
\begin{itemize}
    \item Writing the whole flow field to disk
    \item Computing the energy spectrum
    \item Computing the enstrophy spectrum
    \item Computing various structure functions
    \item Perform visualization using ParaView and Catalyst \cite{ParaViewCatalyst}
\end{itemize}
All of these writers are parallelized over the MPI ranks of the simulation and require minimal additional memory.

\section{Optimization Strategies}

The third-order Runge Kutta scheme employed for time-stepping requires three evaluations of the time derivative $\partial_t \bu^{\Delta}$. The focus of this section will therefore be on the optimization of the kernels used in Algorithm $2$. Furthermore, all the benchmarking will be performed on the GPU partition of the Piz Daint compute cluster \cite{PizDaint}. Each node contains an Intel Xeon E5-2690 v3 CPU paired with a single NVIDIA Tesla P100 GPU. Device memory is limited to 16GB, while the host provides 64GB of memory.

\subsection{Fourier Transform}

As the Fourier transforms in \mbox{Algorithm 2} dominate its computational complexity, they can be expected to contribute to a significant portion of the total computational time. It is therefore crucial to optimize them meticulously. To this end, we choose to use the highly optimized and readily available FFT implementations FFTW \cite{FFTW05} on the host and cuFFT \cite{cuFFT} on the device. Both implementations excel when the size of the Fourier transform only contains small prime factors. For general sizes, slower algorithms must be used. This enables for some quite strong optimization by noticing that the 2/3-law for dealiasing can be relaxed to allow larger padding sizes. As the FFT is only applied to padded data, the size of this data can be chosen arbitrarily as long as it is larger than $\frac{3}{2}N$. By default, we round the padded data's size up to the next value of the form $2^a3^b5^c7^d$ enabling both FFTW and cuFFT to make use of their optimized algorithms. Nonetheless, due to hardware and implementation details, increasing the padding size even more might still result in an overall speedup of the FFT computation. %(see Figure \ref{fig:fftbench}).
On any new system, we therefore once run a benchmark of only the Fourier transform storing its results and extracting the optimal padding size from that generated data. If no suitable size is found in the benchmark results, we choose the next larger $N$ with prime factors at most 7 instead. This strategy guarantees the optimality of the FFT on each system the simulation is deployed on.
% \begin{figure}
%     \includegraphics[width=0.45\textwidth]{img/fft_benchmark.png}
%     \caption{Benchmarking results of FFTW (cpu) and cuFFT (gpu) Fourier transform libraries on both 2D and 3D domains on Piz Daint. The $x$-axis shows the size $N$ of the domain, while the time needed for a single FFT is plotted in the $y$-axis. The domain sizes $N$ are chosen to be of the form $N = 2^a3^b5^c7^d$.}
%     \label{fig:fftbench}
% \end{figure}

Having an optimal FFT implementation on a single node is not yet sufficient for problems where the domain is distributed over multiple compute nodes. Although FFTW provides an MPI implementation for exactly this case, cuFFT can only distribute over multiple GPUs given they are managed by the same process. We are therefore forced to perform the Fourier transform in at least three separate steps. First, we can apply it batchwise along the two axis where each rank owns all the data. After, we transpose the data to align the formerly distributed axis with one of the rank-local ones. This then enables us to apply another batched Fourier transform along this remaining axis. As the node interconnect bandwidth can be assumed to be significantly lower than the GPU memory bandwidth, special care should be taken when optimizing the required transpose operation.

\subsection{Padding}

In a naive implementation, padding the data with zeros would require an expensive \verb|MPI_Alltoallw| call to redistribute data between ranks. However, by merging the padding with the Fourier transform kernel, we can omit this call completely. Note that if the Fourier transform is applied along a certain axis, it is sufficient for the data to be only padded along the same axis. This way, all the communication can be condensed into the single \verb|MPI_Alltoall| call required by the Fourier transform itself. An illustration of the resulting padded Fourier transform kernel is shown in Figure \ref{fig:padding} where the algorithm is decomposed into the following five fundamental steps:
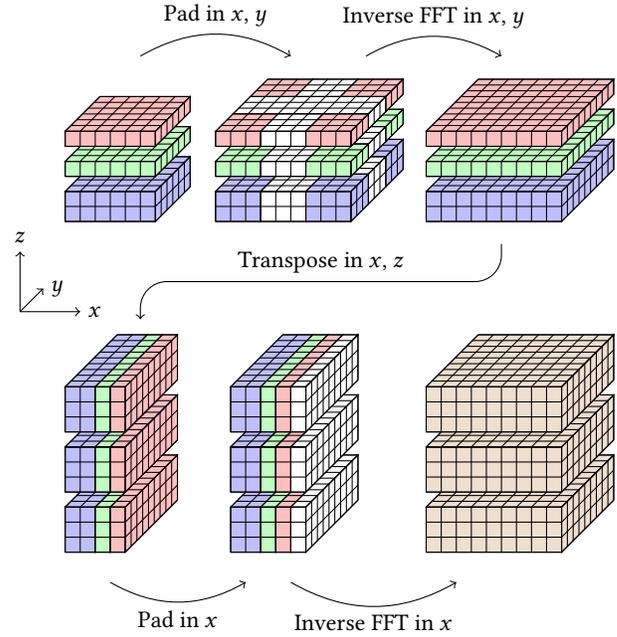
\begin{figure}
    \begin{tikzpicture}
        \def\colorbox(#1,#2,#3,#4,#5,#6,#7,#8) {%
            \draw[fill=#8] (#1, #2, #3) -- (#1+#4*#7, #2, #3) -- (#1+#4*#7, #2+#5*#7, #3) -- (#1, #2+#5*#7, #3) -- cycle;
            \draw[fill=#8] (#1, #2+#5*#7, #3) -- (#1+#4*#7, #2+#5*#7, #3) -- (#1+#4*#7, #2+#5*#7, #3-#6*#7) -- (#1, #2+#5*#7, #3-#6*#7) -- cycle;
            \draw[fill=#8] (#1+#4*#7, #2, #3) -- (#1+#4*#7, #2, #3-#6*#7) -- (#1+#4*#7, #2+#5*#7, #3-#6*#7) -- (#1+#4*#7, #2+#5*#7, #3) -- cycle;
        }
        \def\wirebox(#1,#2,#3,#4,#5,#6,#7) {%
            \foreach \i in {0,...,#4} {
                \draw[] (#1+\i*#7, #2, #3) -- (#1+\i*#7, #2+#5*#7, #3) -- (#1+\i*#7, #2+#5*#7, #3-#6*#7);
            }
            \foreach \i in {0,...,#6} {
                \draw[] (#1, #2+#5*#7, #3-\i*#7) -- (#1+#4*#7, #2+#5*#7, #3-\i*#7) -- (#1+#4*#7, #2, #3-\i*#7);
            }
            \foreach \i in {0,...,#5} {
                \draw[] (#1, #2+\i*#7, #3) -- (#1+#4*#7, #2+\i*#7, #3) -- (#1+#4*#7, #2+\i*#7, #3-#6*#7);
            }
        }
        \def\fullbox(#1,#2,#3,#4,#5,#6,#7,#8) {%
            \colorbox(#1,#2,#3,#4,#5,#6,#7,#8)
            \wirebox(#1,#2,#3,#4,#5,#6,#7)
        }
        
        \def\dx{0.2}

        \fullbox(0,0,0,6,2,6,\dx,white!75!blue)
        \fullbox(0,3*\dx,0,6,1,6,\dx,white!75!green)
        \fullbox(0,5*\dx,0,6,1,6,\dx,white!75!red)

        \draw[->] (5*\dx,11*\dx) to[bend left] node[midway,above] {Pad in $x$, $y$} (15*\dx,11*\dx);

        \colorbox(10*\dx,0,-6*\dx,3,2,3,\dx,white!75!blue)
        \colorbox(13*\dx,0,-6*\dx,3,2,3,\dx,white)
        \colorbox(16*\dx,0,-6*\dx,3,2,3,\dx,white!75!blue)
        \colorbox(10*\dx,0,-3*\dx,9,2,3,\dx,white)
        \colorbox(10*\dx,0,0,3,2,3,\dx,white!75!blue)
        \colorbox(13*\dx,0,0,3,2,3,\dx,white)
        \colorbox(16*\dx,0,0,3,2,3,\dx,white!75!blue)
        \wirebox(10*\dx,0,0,9,2,9,\dx)
        \colorbox(10*\dx,3*\dx,-6*\dx,3,1,3,\dx,white!75!green)
        \colorbox(13*\dx,3*\dx,-6*\dx,3,1,3,\dx,white)
        \colorbox(16*\dx,3*\dx,-6*\dx,3,1,3,\dx,white!75!green)
        \colorbox(10*\dx,3*\dx,-3*\dx,9,1,3,\dx,white)
        \colorbox(10*\dx,3*\dx,0,3,1,3,\dx,white!75!green)
        \colorbox(13*\dx,3*\dx,0,3,1,3,\dx,white)
        \colorbox(16*\dx,3*\dx,0,3,1,3,\dx,white!75!green)
        \wirebox(10*\dx,3*\dx,0,9,1,9,\dx)
        \colorbox(10*\dx,5*\dx,-6*\dx,3,1,3,\dx,white!75!red)
        \colorbox(13*\dx,5*\dx,-6*\dx,3,1,3,\dx,white)
        \colorbox(16*\dx,5*\dx,-6*\dx,3,1,3,\dx,white!75!red)
        \colorbox(10*\dx,5*\dx,-3*\dx,9,1,3,\dx,white)
        \colorbox(10*\dx,5*\dx,0,3,1,3,\dx,white!75!red)
        \colorbox(13*\dx,5*\dx,0,3,1,3,\dx,white)
        \colorbox(16*\dx,5*\dx,0,3,1,3,\dx,white!75!red)
        \wirebox(10*\dx,5*\dx,0,9,1,9,\dx)

        \draw[->] (20*\dx,11*\dx) to[bend left] node[midway,above] {Inverse FFT in $x$, $y$} (29*\dx,11*\dx);

        \fullbox(24*\dx,0,0,9,2,9,\dx,white!75!blue)
        \fullbox(24*\dx,3*\dx,0,9,1,9,\dx,white!75!green)
        \fullbox(24*\dx,5*\dx,0,9,1,9,\dx,white!75!red)

        \draw[->,rounded corners=10] (29*\dx,-1.5*\dx) -- ((29*\dx,-4*\dx) -- node[midway,above] {Transpose in $x$, $z$} (5*\dx,-4*\dx) -- (5*\dx,-6.5*\dx);

        \colorbox(0,-22*\dx,0,2,3,9,\dx,white!75!blue)
        \colorbox(2*\dx,-22*\dx,0,1,3,9,\dx,white!75!green)
        \colorbox(3*\dx,-22*\dx,0,1,3,9,\dx,white!75!red)
        \wirebox(0,-22*\dx,0,4,3,9,\dx)
        \colorbox(0,-18*\dx,0,2,3,9,\dx,white!75!blue)
        \colorbox(2*\dx,-18*\dx,0,1,3,9,\dx,white!75!green)
        \colorbox(3*\dx,-18*\dx,0,1,3,9,\dx,white!75!red)
        \wirebox(0,-18*\dx,0,4,3,9,\dx)
        \colorbox(0,-14*\dx,0,2,3,9,\dx,white!75!blue)
        \colorbox(2*\dx,-14*\dx,0,1,3,9,\dx,white!75!green)
        \colorbox(3*\dx,-14*\dx,0,1,3,9,\dx,white!75!red)
        \wirebox(0,-14*\dx,0,4,3,9,\dx)

        \draw[->] (3*\dx,-24*\dx) to[bend right] node[midway,below] {Pad in $x$} (12*\dx,-24*\dx);

        \colorbox(11*\dx,-22*\dx,0,2,3,9,\dx,white!75!blue)
        \colorbox(13*\dx,-22*\dx,0,1,3,9,\dx,white!75!green)
        \colorbox(14*\dx,-22*\dx,0,1,3,9,\dx,white!75!red)
        \colorbox(15*\dx,-22*\dx,0,1,3,9,\dx,white)
        \wirebox(11*\dx,-22*\dx,0,5,3,9,\dx)
        \colorbox(11*\dx,-18*\dx,0,2,3,9,\dx,white!75!blue)
        \colorbox(13*\dx,-18*\dx,0,1,3,9,\dx,white!75!green)
        \colorbox(14*\dx,-18*\dx,0,1,3,9,\dx,white!75!red)
        \colorbox(15*\dx,-18*\dx,0,1,3,9,\dx,white)
        \wirebox(11*\dx,-18*\dx,0,5,3,9,\dx)
        \colorbox(11*\dx,-14*\dx,0,2,3,9,\dx,white!75!blue)
        \colorbox(13*\dx,-14*\dx,0,1,3,9,\dx,white!75!green)
        \colorbox(14*\dx,-14*\dx,0,1,3,9,\dx,white!75!red)
        \colorbox(15*\dx,-14*\dx,0,1,3,9,\dx,white)
        \wirebox(11*\dx,-14*\dx,0,5,3,9,\dx)

        \draw[->] (15*\dx,-24*\dx) to[bend right] node[midway,below] {Inverse FFT in $x$} (26*\dx,-24*\dx);

        \fullbox(24*\dx,-22*\dx,0,9,3,9,\dx,white!75!brown)
        \fullbox(24*\dx,-18*\dx,0,9,3,9,\dx,white!75!brown)
        \fullbox(24*\dx,-14*\dx,0,9,3,9,\dx,white!75!brown)

        \draw[->] (-3*\dx,-6*\dx,0) -- (1*\dx,-6*\dx,0) node[right] {$x$};
        \draw[->] (-3*\dx,-6*\dx,0) -- (-3*\dx,-6*\dx,-4*\dx) node[right] {$y$};
        \draw[->] (-3*\dx,-6*\dx,0) -- (-3*\dx,-2*\dx,0) node[above] {$z$};
    \end{tikzpicture}
    \caption{Illustration of the combined padding and inverse Fourier transform kernels for a domain size $N = 6$, padded size $N_{pad} = 9$, and $3$ MPI ranks. Each cell of the domain is drawn as a small cube, while the subdomain currently located on a single MPI rank is illustrated as a block of these small cubes. The subdomains located on each rank are spaced apart from each other to reinforce the notion of them not residing in the same shared memory space. The initial data in Fourier space (top left) is complex valued and has shape $N \times N \times \left(\floor{N/2}+1\right)$, while the output of the kernel (bottom right) is real valued and has shape $N_{pad} \times N_{pad} \times N_{pad}$.}
    \label{fig:padding}
\end{figure} \\\\
\noindent {\bf Algorithm $3$:}
\begin{itemize}
    \item[1.] Pad the data with zeros in the rank-local $x$- and $y$-axes.
    \item[2.] Apply a batched complex-to-complex inverse Fourier transform along the $x$- and $y$-axes.
    \item[3.] Transpose by swapping the $x$- and $z$-axes
    \item[4.] Pad with zeros in the new $x$-axis (formerly the $z$-axis)
    \item[5.] Apply a batched complex-to-real inverse Fourier transform along the new $x$-axis.
\end{itemize} \null
Note that in this fused padding and Fourier transform kernel, the amount of data to be transposed is reduced by a factor of approximately $1.5$, as it is only partially padded at that stage. Compared to the unfused kernels, this padded Fourier transform thus reduces the communicated data by almost a factor three. Furthermore, the forward Fourier transform needed in the simulation simultaneously removes the padding from the data. This is achieved by executing Algorithm $3$ in reverse.

\subsection{Transpose}

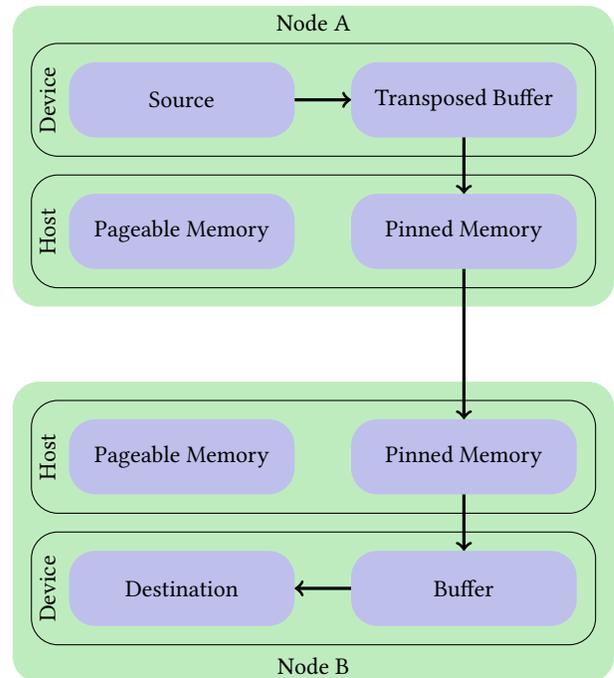
\begin{figure}
    \begin{tikzpicture}
        \def\ax{0}
        \def\ay{5}
        \def\bx{0}
        \def\by{0}
        \def\dx{8}
        \def\dy{4}
        
        \fill[rounded corners=10, fill=black!30!green!25] (\ax,\ay) rectangle (\ax+\dx,\ay+\dy);
        \node[anchor=north] at (\ax+\dx/2,\ay+\dy) {Node A};
        \draw[rounded corners=10] (\ax+0.25,\ay+0.25) rectangle (\ax+\dx-0.25,\ay+\dy/2-0.25);
        \node[anchor=north, rotate=90] at (\ax+0.25,\ay+\dy/4) {Host};
        \fill[rounded corners=10, fill=black!30!blue!25] (\ax+0.75,\ay+0.5) rectangle (\ax+\dx/2-0.25,\ay+\dy/2-0.5);
        \node at (\ax+\dx/4+0.25,\ay+\dy/4) {Pageable Memory};
        \fill[rounded corners=10, fill=black!30!blue!25] (\ax+\dx/2+0.5,\ay+0.5) rectangle (\ax+\dx-0.5,\ay+\dy/2-0.5);
        \node at (\ax+3*\dx/4,\ay+\dy/4) {Pinned Memory};
        \draw[rounded corners=10] (\ax+0.25,\ay+\dy/2) rectangle (\ax+\dx-0.25,\ay+\dy-0.5);
        \node[anchor=north, rotate=90] at (\ax+0.25,\ay+0.75*\dy-0.25) {Device};
        \fill[rounded corners=10, fill=black!30!blue!25] (\ax+0.75,\ay+\dy/2+0.25) rectangle (\ax+\dx/2-0.25,\ay+\dy-0.75);
        \node at (\ax+\dx/4+0.25,\ay+3*\dy/4-0.25) {Source};
        \fill[rounded corners=10, fill=black!30!blue!25] (\ax+\dx/2+0.5,\ay+\dy/2+0.25) rectangle (\ax+\dx-0.5,\ay+\dy-0.75);
        \node at (\ax+3*\dx/4,\ay+3*\dy/4-0.25) {Transposed Buffer};
        
        \fill[rounded corners=10, fill=black!30!green!25] (\bx,\by) rectangle (\bx+\dx,\by+\dy);
        \node[anchor=south] at (\bx+\dx/2,\by) {Node B};
        \draw[rounded corners=10] (\bx+0.25,\by+0.5) rectangle (\bx+\dx-0.25,\by+\dy/2);
        \node[anchor=north, rotate=90] at (\bx+0.25,\by+\dy/4+0.25) {Device};
        \fill[rounded corners=10, fill=black!30!blue!25] (\bx+0.75,\by+0.75) rectangle (\bx+\dx/2-0.25,\by+\dy/2-0.25);
        \node at (\bx+\dx/4+0.25,\by+\dy/4+0.25) {Destination};
        \fill[rounded corners=10, fill=black!30!blue!25] (\bx+\dx/2+0.5,\by+0.75) rectangle (\bx+\dx-0.5,\by+\dy/2-0.25);
        \node at (\bx+3*\dx/4,\by+\dy/4+0.25) {Buffer};
        \draw[rounded corners=10] (\bx+0.25,\by+\dy/2+0.25) rectangle (\bx+\dx-0.25,\by+\dy-0.25);
        \node[anchor=north, rotate=90] at (\bx+0.25,\by+0.75*\dy) {Host};
        \fill[rounded corners=10, fill=black!30!blue!25] (\bx+0.75,\by+\dy/2+0.5) rectangle (\bx+\dx/2-0.25,\by+\dy-0.5);
        \node at (\bx+\dx/4+0.25,\bx+3*\dy/4) {Pageable Memory};
        \fill[rounded corners=10, fill=black!30!blue!25] (\bx+\dx/2+0.5,\by+\dy/2+0.5) rectangle (\bx+\dx-0.5,\by+\dy-0.5);
        \node at (\bx+3*\dx/4,\bx+3*\dy/4) {Pinned Memory};

        \draw[->,very thick] (\ax+\dx/2-0.25,\ay+3*\dy/4-0.25) -- (\ax+\dx/2+0.5,\ay+3*\dy/4-0.25);
        \draw[->,very thick] (\ax+3*\dx/4,\ay+\dy/2+0.25) -- (\ax+3*\dx/4,\ay+\dy/2-0.5);
        \draw[->,very thick] (\ax+3*\dx/4,\ay+0.5) -- (\bx+3*\dx/4,\by+\dy-0.5);
        \draw[->,very thick] (\bx+3*\dx/4,\by+\dy/2+0.5) -- (\bx+3*\dx/4,\by+\dy/2-0.25);
        \draw[->,very thick] (\bx+\dx/2+0.5,\by+\dy/4+0.25) -- (\bx+\dx/2-0.25,\by+\dy/4+0.25);
    \end{tikzpicture}
    \caption{Unidirectional data path between two nodes for the distributed transpose operation on GPUs. We first preprocess the data by transposing it locally on the GPU. This data is then copied down into pinned memory on the host bypassing pageable memory doubling the bandwidth for device to host transfers. This is followed by sending the data asynchronously over MPI to the receiving node. There, the data is received directly into another pinned memory block and consequently copied onto the GPU. Finally, the data is then distributed into the destination array containing the transposed data.}
    \label{fig:distributed_transpose}
\end{figure}

The transpose operation is the kernel with the most potential for optimization as it contains slow MPI communication between compute nodes and a strided memory access pattern resulting in cache unfriendly code and diminishing of the memory bandwidth. During the distributed transpose operation, the data is redistributed among the MPI ranks according to step 3 depicted in Figure \ref{fig:padding}. Each rank splits its data along the $x$-axis where the sizes of these blocks are given by their respective shapes in the transposed data. The slices are then transposed locally and sent to their corresponding ranks. This redistribution of the data requires copying the data from the GPU to the host, sending it to the other MPI ranks with an \verb|MPI_Alltoall| call, and finally copying the received data back onto the GPU while at some point also performing the local transpose. The path a single packet of data takes in the implementation of the transpose is visualized in Figure \ref{fig:distributed_transpose}. The local transpose operation is performed on the GPU benefiting from its higher bandwidth compared to the CPU. The data is then copied down into page locked memory enabling asynchronous transfers and doubling the bandwidth. From there it is sent to the receiver node on which it takes the same path in reverse. All of these operations are independent between the blocks sent to different nodes. It is therefore possible to completely parallelize them by overlapping most of the local transposes and moving data from and to the GPU with the slow inter node communication. However, note that communication can only start if the first block of memory was already transposed and copied to host memory. Hence, the first block is not able to be overlapped with the communication making it necessary to also optimize the locally performed operations. The local transpose is a prime candidate for optimization, as a naive implementation has very bad cache locality. This can be solved with the well known technique called blocking \cite{Drepper2007} where the transpose of an array is evaluated in small blocklets each fully fitting into cache. This way, no data that will later still be used is ever displaced from the cache allowing the algorithm to fully exploit the memory bandwidth.

\subsection{Memory Optimizations}

Computing the Fourier transform over data shared by multiple MPI ranks is a lot more expensive than computing a rank local Fourier transform due to the transpose operation required. We therefore want to fit large domains onto a single compute node. The main restriction for this is the amount of memory needed by the simulation. Reducing the memory requirements for a single kernel is difficult if not impossible. However, all the kernels are called sequentially and none of them require storing a significant amount of information across iterations of the simulation. Because most kernels are very cheap frequent allocation and deallocation of memory would result in a considerable hit on performance. We therefore implemented the reduction of memory requirements by introducing the concept of shared \verb|Workspaces|. A workspace is nothing but a contiguous buffer of memory that is provided to a kernel as storage for in- or outputs or temporary results. Because these workspaces are managed by the simulation itself and not by each individual kernel, they can be shared across kernel boundaries enabling a significant portion of the buffers to overlap. This resulted in 12.5\% memory savings for the single node simulation and in over 50\% savings for the MPI implementation.

% \subsubsection{Blocking}

% \subsubsection{Pinned Memory}

% \subsubsection{Asynchronous Communication}

\section{Validation}

For the validation of the code, over 150 tests have been implemented. First, the 2D solver is validated by simulating some analytically known solutions to the incompressible Euler equations and checking their convergence. Examples thereof are the 2D Taylor-Green Vortex \cite{TaylorGreen1937}, or the vortex patch \cite{ParesMishra2020}. The 3D solver is then checked by initializing the flow field with the previously checked 2D initial conditions while keeping the extra dimension constant. Both the 2D and the 3D versions are additionally verified visually by computing non-stationary solutions which were simulated/observed in previous work such as the discontinuous shear layer in \cite{LMP1}, the double shear layer from \cite{BROWN1995165}, or the 3D Taylor-Green vortex \cite{TaylorGreen1937}. This is especially important for the 3D incompressible Euler code, as there are no non-trivial stationary solutions in that case. Finally, we compute observables and the structure functions of some statistical solutions and compare them with structure functions from previous experiments done with different solvers as in \cite{LMP1,CPP1}.

\section{Performance Assessment}

Denote the host memory bandwidth by $\beta^H$, the device memory bandwidth by $\beta^D$, the bandwidth of the link between host and device by $\beta^{D-H}$, and the MPI communication bandwidth by $\beta^{MPI}$. Their values as measured on Piz Daint can be found in Table \ref{tab:bandwidth}.
\begin{table}
    \caption{Measured Bandwidths on the Piz Daint Cluster used to compute the lower bounds on the runtime of the compute kernels.}
    \label{tab:bandwidth}
    \begin{tabular}{cc} \toprule
        Type & Speed \\ \midrule
        $\beta^{H}$ & 21.24GB/s \\
        $\beta^{D}$ & 488.64GB/s \\
        $\beta^{H-D}$ & 19.24GB/s \\
        $\beta^{MPI}$ & 9GB/s \\ \bottomrule
    \end{tabular}
\end{table}
Henceforth, we will assume that the FFTW and cuFFT implementations are optimal and not include them in the performance assessment of the simulation. The execution speed of all other kernels including the transpose operation will be compared against their theoretical optimum obtained by carefully analyzing their required data movements. We denote the resolution of the simulation domain by $N$, the padded size by $N_{pad}$, and the number of MPI processes by $p$. Additionally, $s$ will denote the size of the underlying scalar datatype of our simulation in bytes. This will either be $s = 4$ for floats or $s = 8$ for doubles.

\subsection{Padding and Unpadding}

Padding is performed in two steps as depicted in Figure \ref{fig:padding}.  As both of these padding operations only need to copy data to another array without performing any computations on them, the kernel is certainly memory bound. Hence, we only need to take the memory bandwidth of the system into account in order to obtain a lower bound for the runtime. Both steps must read each element of the input array exactly once and write to each element of the output exactly once as well. Counting the total number of (complex valued) elements in the arrays, we obtain $3N^2\frac{\floor{N/2}+1}{p}$, $3N_{pad}^2\frac{\floor{N/2}+1}{p}$, $3(\floor{N/2}+1)N_{pad}\frac{N_{pad}}{p}$, and $3(\floor{N_{pad}/2}+1)N_{pad}\frac{N_{pad}}{p}$. Collecting all the reads and writes on these elements and combining them with the memory bandwidth of our system, we obtain the following lower bound for the execution time of the padding kernel:
\begin{equation}
    t^{pad}(N, p) \approx \frac{213}{16}\frac{N^3}{p}\frac{2s}{\beta}
\end{equation}
where we have used that $\floor{N/2}+1 \approx N/2$ and $N_{pad} \approx \frac{3}{2}N$ for large $N$. $\beta$ takes the value of either the host memory bandwidth $\beta^H$ or the device memory bandwidth $\beta^D$ depending on where the kernel is executed.

Although the unpadding operation is very similar, there are still a few key differences compared to the padding operation. Firstly, as unpadding is performed on the upper triangular matrix $B = u \otimes u$, we have $6$ instead of $3$ components to unpad. Secondly, the unpadding does not require reading the high frequency modes, as they are discarded anyway. Considering these changes, the lower bound for the execution time of the unpadding operation is given by
\begin{equation}
    t^{unpad}(N, p) \approx \frac{78}{4}\frac{N^3}{p}\frac{2s}{\beta}
\end{equation}
where we have used the same simplifications as above.

Table \ref{tab:rt} shows the timings and achieved performances of our kernels. Note that the padding kernel achieves around 50\% of the optimal performance we computed. This is most probably due to expensive index computations and the branching operation to decide whether to copy or write zero to the destination array. This hypothesis is also supported by the fact that the unpadding operation achieves almost 80\% of the optimal performance, as it can reuse a single index computation for 6 instead of 3 components of the array. Additionally, it does not require branching leading to its improved performance results.

\subsection{Transpose}

In order to find the optimal execution speed to the transpose kernel, we need to know the critical path of the algorithm limiting the speed of the solver. As the pre-, and post-processing, and communication are independent, they can be overlapped in order to maximize parallelization. The algorithm must start by locally transposing a single block of data containing $3\frac{N_{pad}}{p}N_{pad}\frac{\floor{N/2}+1}{p}$ complex valued elements on the GPU. This is followed by copying the block of data down into host memory, while simultaneously starting the local transpose of the second block. After the whole first block is located in host memory, the MPI communication can begin and the last block of data will be communicated after each MPI rank has sent its $3N_{pad}N_{pad}\frac{\floor{N/2}+1}{p}$ local elements. The algorithm then finishes by copying this last block of data from the host onto the GPU and finally inserting that data in the destination buffer. Adding up all of these contributions, the lower bound for the execution time of the transpose operation of a single velocity component is given by
\begin{equation}
    t^{T}(N, p) \approx \frac{9}{2}\frac{N^3}{p^2}\frac{2s}{\beta^D} + \frac{9}{4}\frac{N^3}{p^2}\frac{2s}{\beta^{H-D}} + \frac{9}{8}\frac{N^3}{p}\frac{2s}{\beta^{MPI}}.
\end{equation}
To find the times for transposing $u$ or $B$, this estimate can simply be multiplied by their respective number of components.

The benchmarking results in Table \ref{tab:rt} show that the kernels achieve around 50\% of the previously computed optimal performance estimate. This changes, however, if $\beta^{MPI}$ is changed from the point-to-point message bandwidth to the bandwidth of an all-to-all communication reducing it to approximately $\beta^{MPI} =$ 4GB/s. This has a large effect on the value of $t^{opt}$ increasing it by almost a factor of two. Consequently also the efficiency is increased by the same amount. The resulting timings and efficiencies are listed in Table \ref{tab:rt} in parentheses. Note that the transpose of $B$ achieves over 100\% efficiency. This can be contributed to the new upper bound of the MPI bandwidth obtained through a benchmark of \verb|MPI_Alltoall|. Our custom implementation beats the native MPI on Piz Daint for messages larger than 1MB, achieving a transfer speed of approximately 5GB/s instead of the 4GB/s of the native MPI in the case covered by the benchmark problem considered.

\begin{table*}
    \caption{Runtimes of the compute kernels for a single precision floating point simulation of resolution $N = 512$ computed on the GPUs of $8$ MPI ranks on the Piz Daint cluster. We measure the runtime $t$ of a single call to the kernel, compare it to the lower bound $t^{opt}$ previously computed, and also give the contribution to the total runtime of the simulation for each kernel. The execution time $t$, and the efficiency $t^{opt}/t$ for the transpose is given twice. The value in parentheses is obtained by taking the MPI\_Alltoall bandwidth instead of the point-to-point bandwidth on Piz Daint for the computation of the lower bound $t^{opt}$.}
    \label{tab:rt}
    \begin{tabular}{ccccc} \toprule
        Kernel & $t$ & $t^{opt}$ & $t^{opt}/t$ & Total Runtime \% \\ \midrule
        Padding & 6.982ms & 3.657ms & 52.4\% & 1.7\% \\
        Unpadding & 6.783ms & 5.356ms & 79\% & 3\% \\
        Transpose $u$ & 120.765ms & 56.681ms (119.596ms) & 46.9\% (99\%) & 29.8\% \\
        Transpose $B$ & 200.895ms & 113.362ms (239.191ms) & 56.4\% (119.1\%) & 49.6\% \\ \bottomrule
    \end{tabular}
\end{table*}

\subsection{Parallel Scaling}

\begin{figure}
    \begin{subfigure}{0.225\textwidth}
        \includegraphics[width=\textwidth]{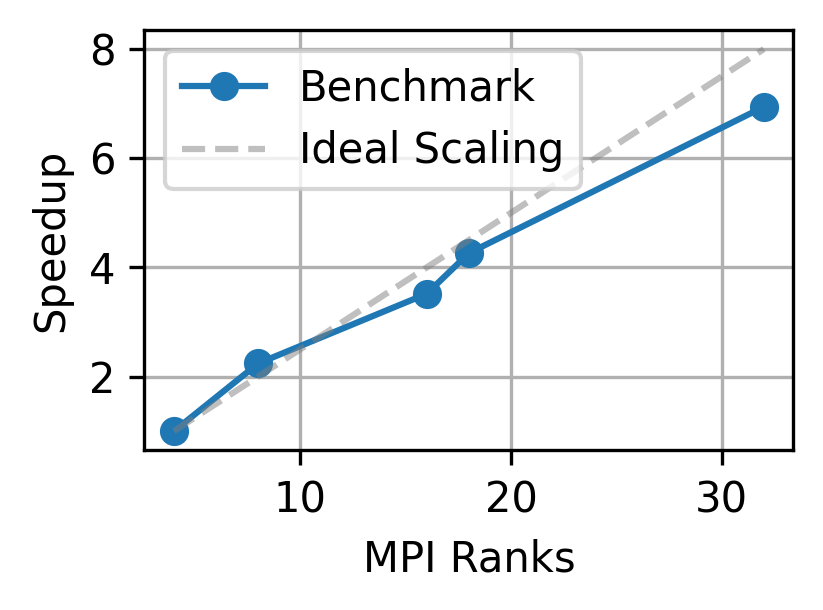}
    \end{subfigure}
    \begin{subfigure}{0.225\textwidth}
        \includegraphics[width=\textwidth]{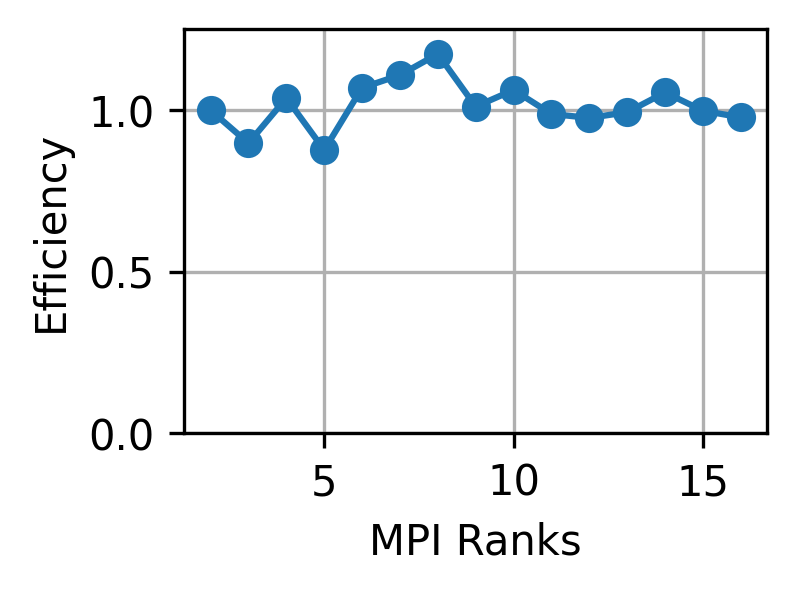}
    \end{subfigure}
    \caption{Strong (left) and weak (right) scaling of our simulation when parallelizing over a single sample. Strong scaling is computed at a domain size of $N = 512$ and $p = 4$ MPI ranks in the base case. Scaling can be observed to be almost optimal. To measure the weak scaling, we start with $N = 256$ parallelized over $p = 2$ MPI ranks. Then, to keep the amount of work per rank constant according to the computational complexity $\cO(N^3\log_2(N))$ of the solver, we add more benchmarks for $p$ up to $16$ ranks while adjusting the domain size $N$ accordingly. We observe the parallel efficiency to be optimal with some configurations even reaching efficiencies over 100\%. This can be attributed to the more efficient overlap of computation with communication due to the smaller size and higher amount of data blocks to be communicated.}
    \label{fig:scaling_N}
\end{figure}

\begin{figure}
    \begin{subfigure}{0.225\textwidth}
        \includegraphics[width=\textwidth]{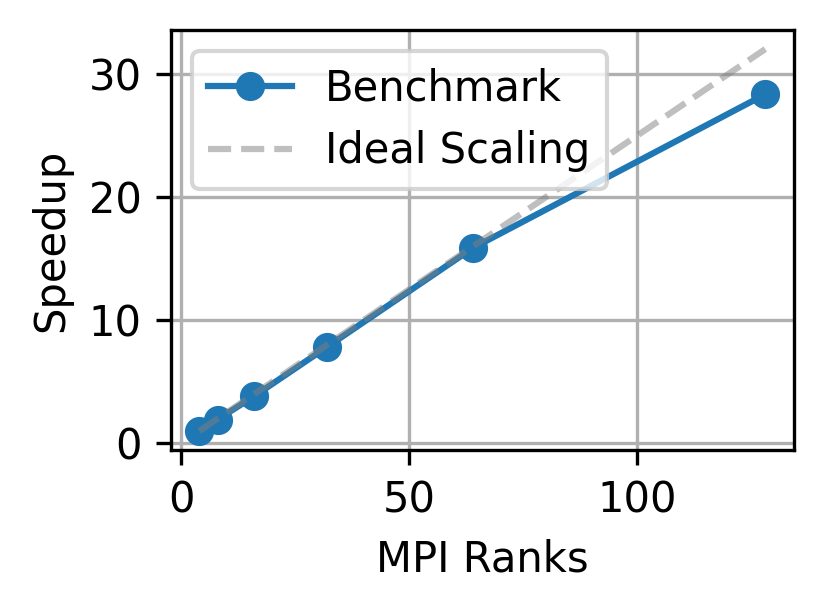}
    \end{subfigure}
    \begin{subfigure}{0.225\textwidth}
        \includegraphics[width=\textwidth]{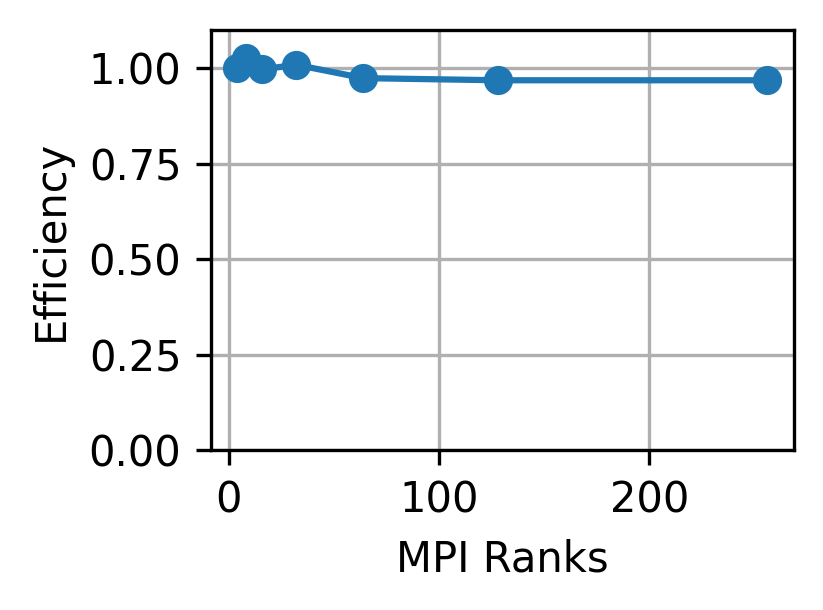}
    \end{subfigure}
    \caption{Strong (left) and weak (right) scaling of our simulation when parallelizing over multiple Monte Carlo samples at a fixed resolution $N = 512$ with 4 MPI ranks per sample. Strong scaling considers $M = 128$ samples, while for the weak scaling, we use $M = 128$ samples computed with $p = 256$ MPI ranks and scale down both $M$ and $p$ down accordingly to keep the work per rank constant. As expected from Monte Carlo sampling, both the strong and the weak scaling are close to optimal, with the only inefficiencies probably coming from parallel accesses to the file system.}
    \label{fig:scaling_M}
\end{figure}

We measure strong and weak scaling of parallelization in both a single sample and over multiple Monte Carlo samples. The results can be found in Figures \ref{fig:scaling_N}, and \ref{fig:scaling_M}. Note that for statistical solutions, we are particularly interested in computing many Monte Carlo samples, even at the expense of some drop in spatial resolution. For a single sample, we therefore usually use close to the fewest number of ranks such that the sample still fits in memory and the main parallelization is done over Monte Carlo samples. This strategy is in agreement with the scaling of the solver, as the strong scaling seems to be slightly better in the number of samples per MPI rank than in the number of MPI ranks per sample. Also, the weak scaling efficiency of both the parallelization of a single sample and parallelization over Monte Carlo samples are almost optimal. The weak scaling efficiency over a single sample (Figure \ref{fig:scaling_N}, right) even grows to be larger than 100\% when using only slightly more ranks than necessary for the problem. Same can be observed for the weak scaling over Monte Carlo samples. After an early peak in efficiency, it levels off at approximately 97\%. This small decrease in efficiency can most probably be attributed to more parallel accesses to the file system.


\subsection{Baselines}

In order to provide a benchmark demonstrating the speed of the whole parallelized FFT implementation, we present a comparison to the FFT used in the HACC (Hardware/Hybrid Accelerated Cosmology Code) introduced in \cite{HABIB201649}. The comparison of our implementation versus the SWFFT used in the HACC can be found in Table \ref{tab:fft}. The benchmark problems are chosen to be representative of the conditions encountered during typical use of our simulation. The domain sizes $N = 768^3$ and $N = 1536^3$ reflect the most common resolutions used in the context of statistical solutions computed with our solver and, due to padding according to the 2/3-dealiasing rule, correspond to an effective resolution of $N = 512^3$ and $N = 1024^3$. Simulations with resolutions lower than that fit onto a single GPU on Piz Daint and therefore use cuFFT directly, while resolutions larger than the $N = 1024^3$ are computationally too expensive to draw the necessary number of Monte Carlo samples from. As SWFFT is highly configurable, we run a grid search over all configurations selecting the fastest one for each individual problem. Nonetheless, our implementation consistently outperforms the SWFFT implementation by approximately 20\%. This is mainly enabled by the fact that for statistical solutions we are able to trade resolution for the number of Monte Carlo samples drawn. The FFT used in our solver can exploit this by optimizing specifically for these (in comparison to HACC) low resolutions. In particular, we are able to use a slab instead of a pencil decomposition resulting in a significant reduction in communication overhead. Furthermore, smaller optimizations like overlapping communication with computation are able to be specifically tuned for the resolutions encountered when computing statistical solutions.
\begin{table}
    \caption{Runtimes of the GPU-based forward and backward Fourier transform performed with SWFFT and our implementation averaged over 8 runs. Domain sizes of $N = 768^3$ and $N = 1536^3$ were tested as these are the most common use cases in our simulation. Furthermore, each domain size was parallelized over $p=8,16,32,64$ MPI ranks. For SWFFT we ran a grid search to find the fastest configuration for each problem instance separately. Nonetheless, our implementation consistently outperformed SWFFT by around 20\%.}
    \label{tab:fft}
    \begin{tabular}{ccccc} \toprule
         $N$ & $p$ & SWFFT   & Ours    & Speedup \\ \midrule
         768 &   8 & 0.9754s & 0.7945s & 22.8\%  \\
         768 &  16 & 0.4994s & 0.4048s & 23.4\%  \\
         768 &  32 &  0.257s & 0.2177s & 18.1\%  \\
         768 &  64 & 0.1382s & 0.1301s &  6.2\%  \\
        1536 &   8 & crashed & 6.5391s & ---     \\
        1536 &  16 &  4.057s & 3.3237s & 22.1\%  \\
        1536 &  32 &   2.06s & 1.7186s & 19.9\%  \\
        1536 &  64 & 1.0646s &   0.87s & 22.4\%  \\ \bottomrule
    \end{tabular}
\end{table}

\section{Application to Turbulent Flows}

Recall that the concept of statistical solutions was originally introduced to restore convergence of the simulation results under mesh refinement. In this section we provide two examples to demonstrate this property. To this end, we use our solver to approximate statistical solutions to given initial conditions and demonstrate convergence of key statistical quantities such as the mean and the variance, as well as convergence in the Wasserstein distance of the empirical measures to a high-resolution reference solution.

\subsection{Taylor-Green Vortex}

The famous Taylor-Green Vortex in 3D \cite{TaylorGreen1937} initializes the flow field with
\begin{equation}
    \begin{aligned}
        u_0(x, y, z) &= A \cos(2\pi x) \sin(2\pi y) \sin(2\pi z) \\
        v_0(x, y, z) &= B \sin(2\pi x) \cos(2\pi y) \sin(2\pi z) \\
        w_0(x, y, z) &= C \sin(2\pi x) \sin(2\pi y) \cos(2\pi z).
    \end{aligned}
\end{equation}
In order for the flow to be divergence free, we need to satisfy the constraint $A + B + C = 0$. To do so, we choose $A = 1$, $B = -1$, and $C = 0$. We add a small perturbation to the initial conditions which is given by the first-order harmonics with random amplitudes.We define $8d$ i.i.d uniformly distributed random variables $\delta_{d,i,j,k} \sim \mathcal{U}_{\left[ -0.025, 0.025 \right]}$. We then define the perturbation $\varepsilon_d(x, y, z)$ on the $d$-th velocity component as
\begin{equation}
    \begin{split}
        &\varepsilon_d(x, y, z) = \frac{1}{8} \sum_{(i,j,k) \in \{0,1\}^3} \delta_{d,i,j,k} \alpha_i(4\pi x) \alpha_j(4\pi y) \alpha_k(4\pi z),\\
        &\mbox{ where } \alpha_i(x) = \begin{cases}
            \sin(x) &\mbox{ if } i = 0, \\
            \cos(x) &\mbox{ if } i = 1
        \end{cases}.
    \end{split}
\end{equation}
Finally, we arrive at the initial measure by perturbing the initial conditions $u_0$, $v_0$, and $w_0$ with $\varepsilon_d$
\begin{equation}
    \begin{aligned}
        u_0(x, y, z) &= \cos(2\pi x) \sin(2\pi y) \sin(2\pi z) + \varepsilon_0(x, y, z) \\
        v_0(x, y, z) &= -\sin(2\pi x) \cos(2\pi y) \sin(2\pi z) + \varepsilon_1(x, y, z) \\
        w_0(x, y, z) &= \varepsilon_2(x, y, z).
    \end{aligned}
\end{equation}

\begin{figure}
    \begin{subfigure}{0.225\textwidth}
        \includegraphics[width=\textwidth]{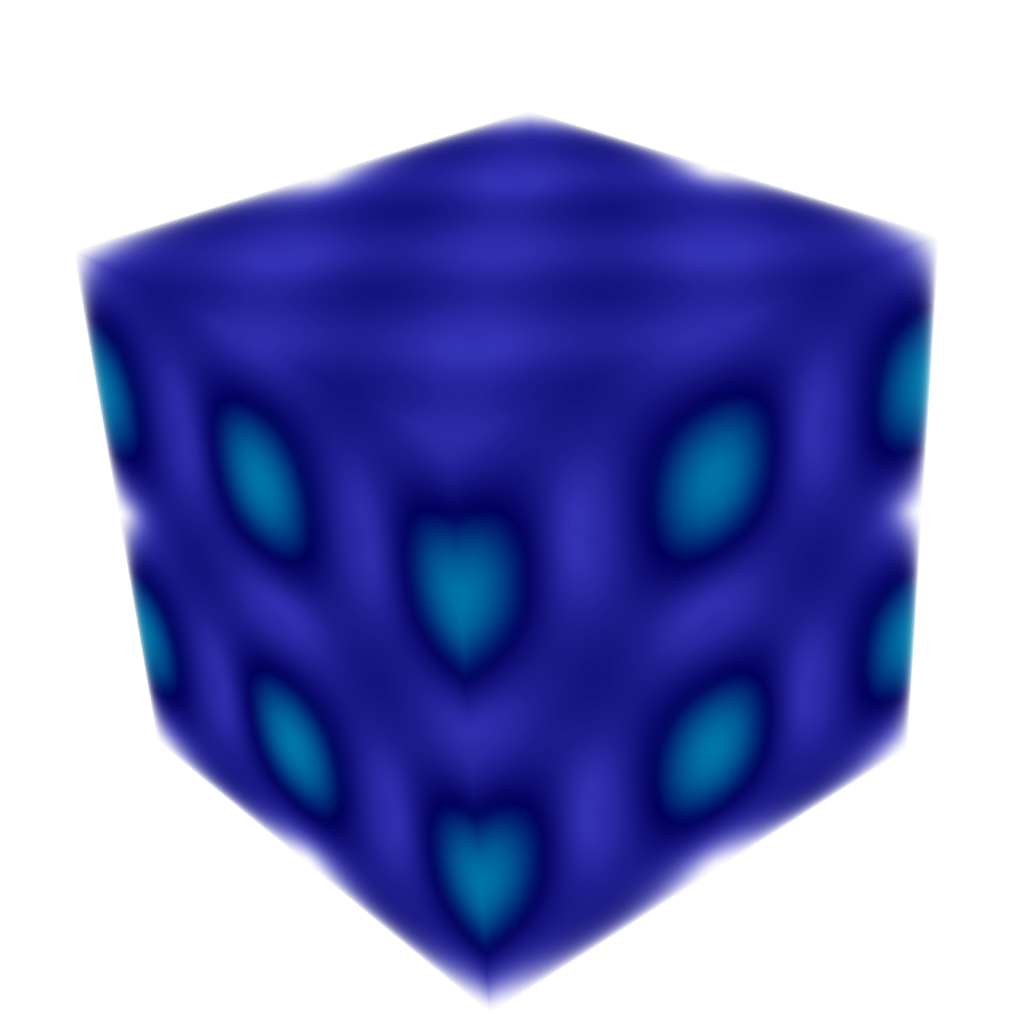}
    \end{subfigure}
    \begin{subfigure}{0.225\textwidth}
        \includegraphics[width=\textwidth]{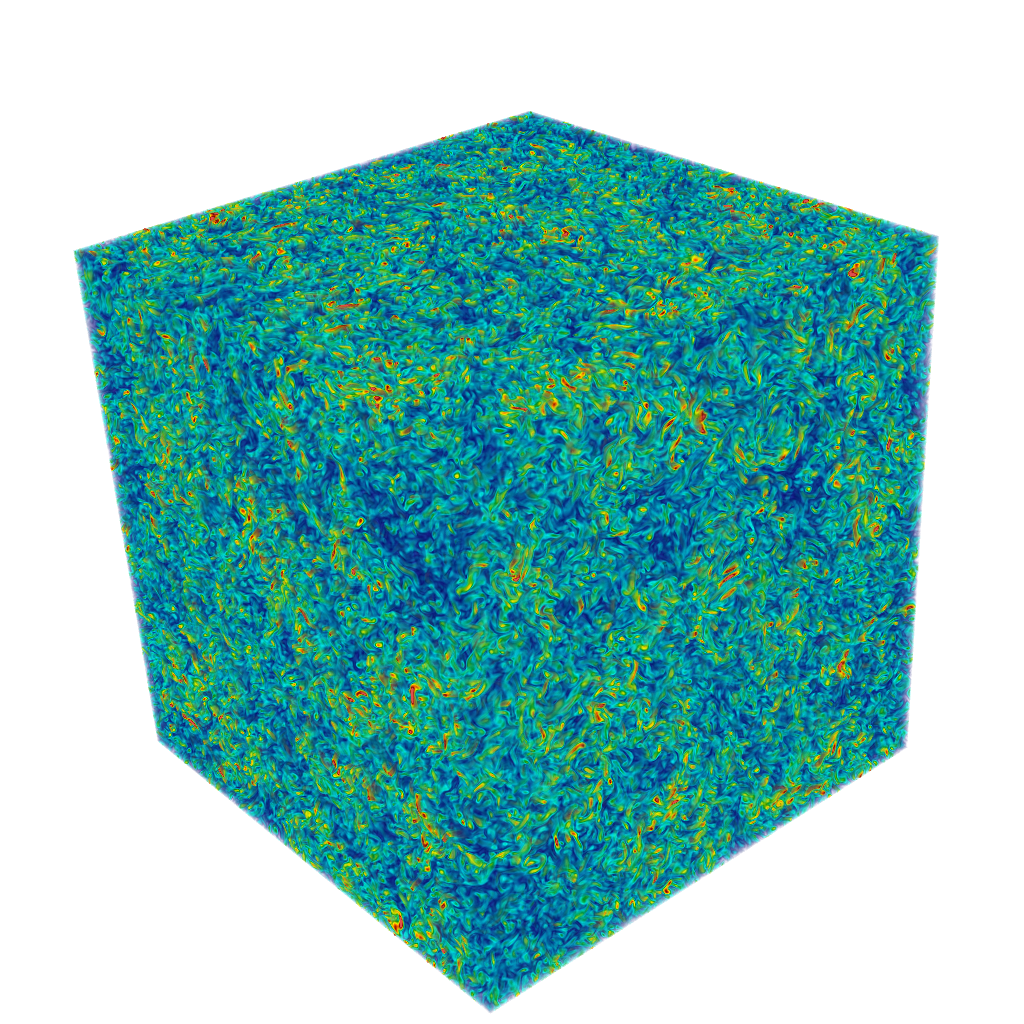}
    \end{subfigure}
    \caption{Vorticity magnitude of the perturbed Taylor-Green vortex at time $t = 0$ (left) and time $t = 5$ (right) at a resolution of $N = 512$.}
    \label{fig:tg_sample}
\end{figure}

For the simulations we choose the hyperviscosity parameter $s = 1.5$ as this increases the range of the intermittent scales in the simulation. A realization of a simulation with the given random initial conditions can be found in Figure \ref{fig:tg_sample}. Upon zooming in, an intricate web of vortex filaments can be seen at time $t = 5$.

\begin{figure}
    \includegraphics[width=0.4\textwidth]{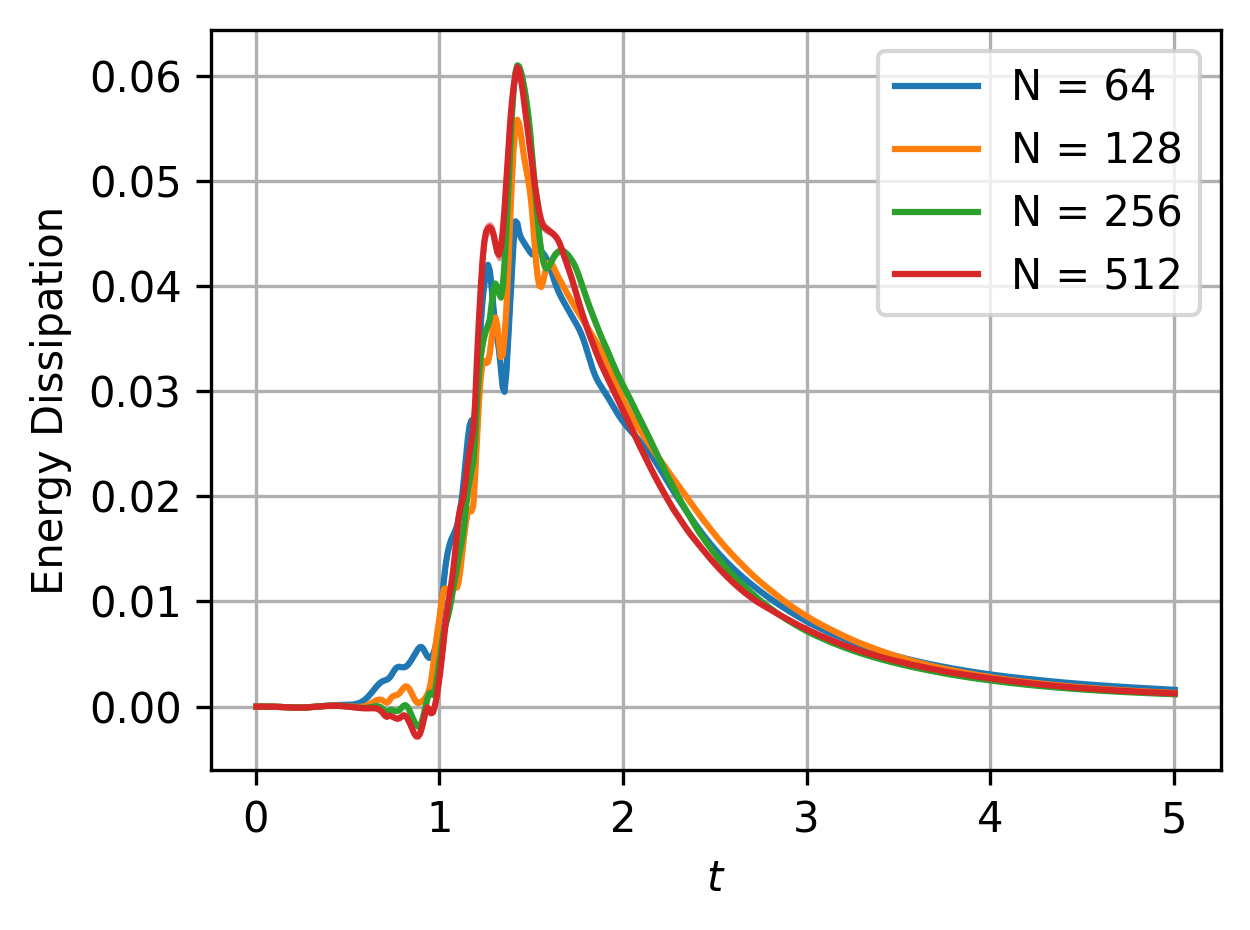}
    \caption{Evolution of energy dissipation $-\frac{\mathrm{d}E}{\mathrm{d}t}$ in the Taylor-Green vortex experiment. Visually, there seems to be a good agreement with energy dissipation rates commonly found in literature \cite{FKMW}.}
    \label{fig:tg_dE}
\end{figure}

According to the famous K41 theory \cite{Frisch1995} we expect to observe an anomalous energy dissipation even in the limit $\nu \to 0$. The flexible IO design of our simulator enables the periodic computation of the kinetic energy of the system at short time intervals. By consequently doing finite differences in time, we are able to compute the evolution of the energy dissipation rate of the system (Figure \ref{fig:tg_dE}). The obtained rates agree well with results in common literature \cite{FKMW} providing additional verification of the solver.

\begin{figure}
    \begin{subfigure}{0.225\textwidth}
        \includegraphics[width=\textwidth]{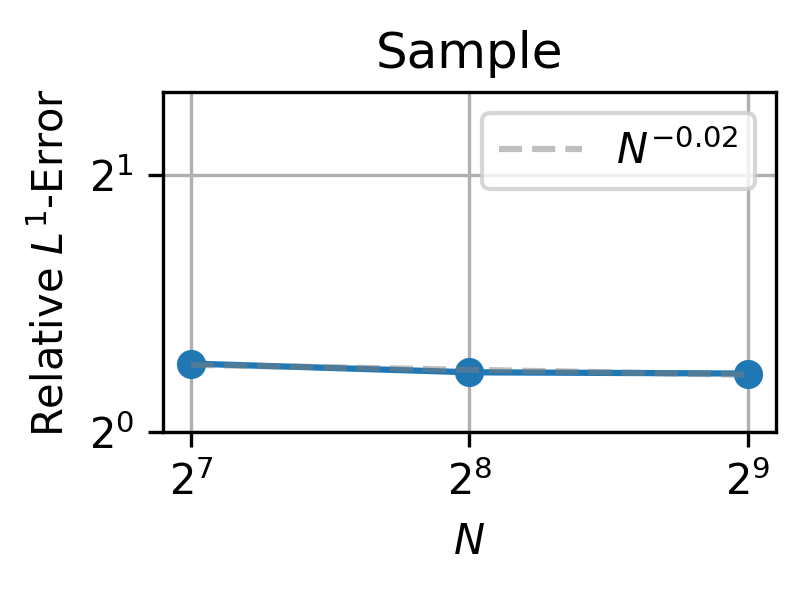}
    \end{subfigure}
    \begin{subfigure}{0.225\textwidth}
        \includegraphics[width=\textwidth]{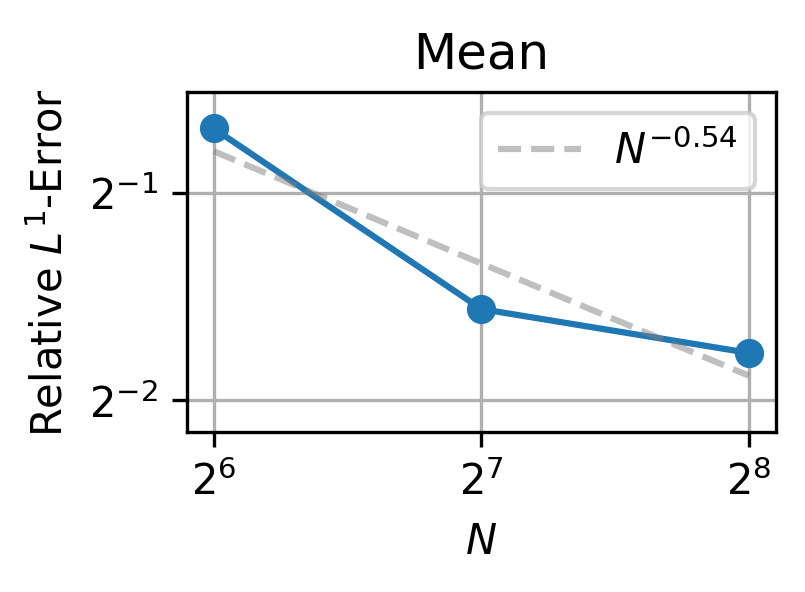}
    \end{subfigure}
    \begin{subfigure}{0.225\textwidth}
        \includegraphics[width=\textwidth]{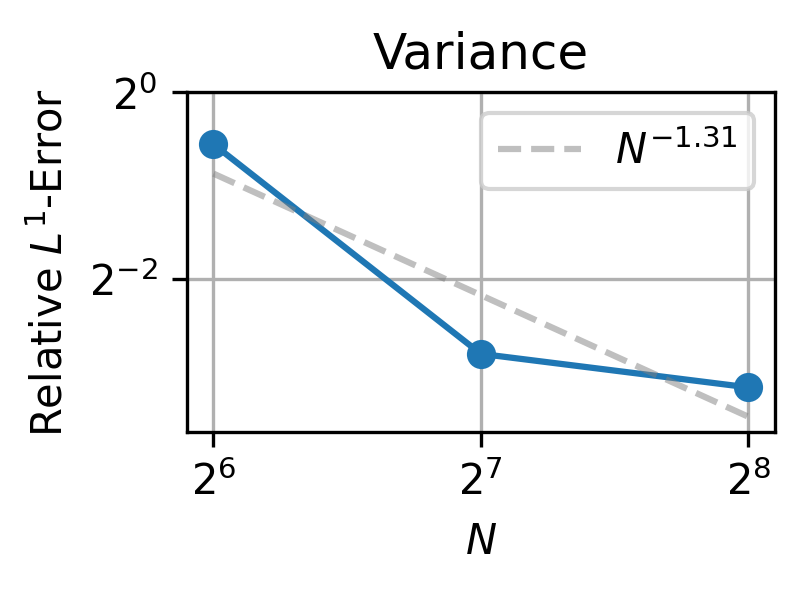}
    \end{subfigure}
    \begin{subfigure}{0.225\textwidth}
        \includegraphics[width=\textwidth]{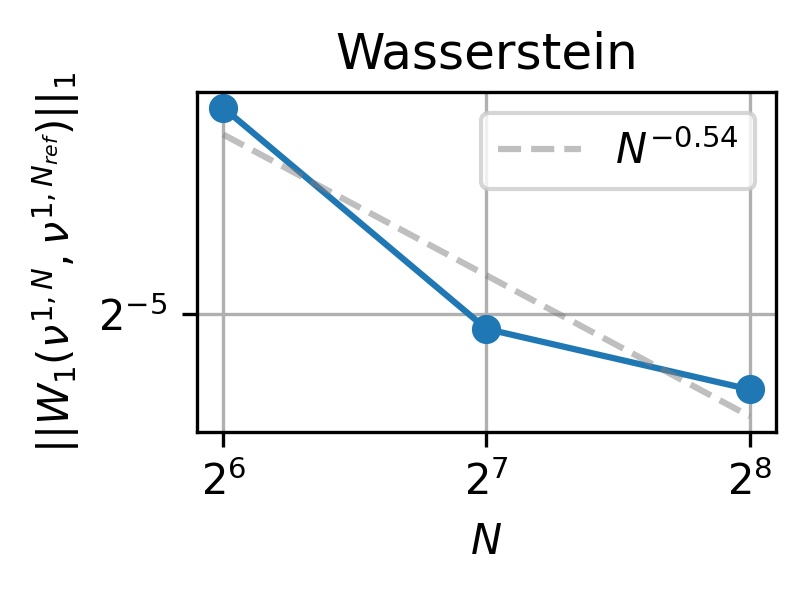}
    \end{subfigure}
    \caption{Convergence rates under mesh refinement of a single sample (top left), the mean (top right), the variance (bottom left), and the 1-point Wasserstein distance (bottom right) for the Taylor-Green Vortex. As expected, no sample wise convergence can be observed which can be explained by the ever smaller vortices being resolved and influencing the larger scales of the flow. Nevertheless, the pointwise mean and variance both seem to converge with a reasonable rate. This is confirmed by the convergence in the Wasserstein distance of the one-point correlation marginals of the statistical solution.}
    \label{fig:tg_convergnce}
\end{figure}

To demonstrate the need for statistical instead of classical solutions, we plot the convergence under mesh refinement of a single sample, the mean, the variance, and the Wasserstein distance in Figure \ref{fig:tg_convergnce}. As expected, we observe no sample-wise convergence. Contrary to this, the statistical quantities considered do converge at a reasonable rate. In particular does the convergence in Wasserstein distance of the one-point correlation marginals imply the convergence of all pointwise statistical moments.

\subsection{Cylindrical Shear Flow}

The Cylindrical Shear Flow is heavily inspired by the Flat Vortex Sheet experiment in \cite{LMP1} and is introduced as a 3D equivalent to the latter. The initial conditions are given by
\begin{equation}
    \begin{aligned}
        u_0(x, y, z) &= \tanh\left(2\pi\frac{r-0.25}{\rho}\right) \\
        v_0(x, y, z) &= 0 \\
        w_0(x, y, z) &= 0
    \end{aligned}
\end{equation}
where $r^2 = (y - 0.5 + \sigma_{\delta}^y(x))^2 + (z - 0.5 + \sigma_{\delta}^z(x))^2$ and $\rho$ is the smoothness parameter. We define the perturbations $\sigma_{\delta}^y(x)$ and $\sigma_{\delta}^z(x)$ in the following way: Let $\alpha_k^y$ and $\alpha_k^z$ be i.i.d uniformly distributed on $[0, 1]$ and let $\beta_k^y$ and $\beta_k^z$ be i.i.d uniformly distributed on $[0, 2\pi]$. Then $\sigma_{\delta}^y(x)$ and $\sigma_{\delta}^z$ are given by
\begin{equation} \label{eq:dst_pert}
    \begin{aligned}
        \sigma_{\delta}^y(x) &= \delta \sum_{k=1}^p \alpha_k^y \sin(2\pi kx - \beta_k^y) \\
        \sigma_{\delta}^z(x) &= \delta \sum_{k=1}^p \alpha_k^z \sin(2\pi kx - \beta_k^z)
    \end{aligned}
\end{equation}
where we have chosen $p = 10$ and $\delta = 0.025$. These initial conditions are well defined in the limit $\rho \to 0$ where the interface between the flow directions becomes discontinuous and are then equal to
\begin{equation}
    \begin{aligned}
        u_0(x, y, z) &= \begin{cases}
            -1 &\mbox{ for } r \leq 0.25 \\
            1 &\mbox{ otherwise}
        \end{cases} \\
        v_0(x, y, z) &= 0 \\
        w_0(x, y, z) &= 0
    \end{aligned}
\end{equation}
where $r$ is defined as above.

The solutions to this experiment contain a multitude of different flow regimes as the turbulence starts to develop at the interface $r = \sqrt{y^2+z^2} \approx 0.25$ and slowly propagates outward.

\begin{figure*}
    \begin{subfigure}{0.45\textwidth}
        \includegraphics[width=\textwidth,trim={100 100 100 100},clip]{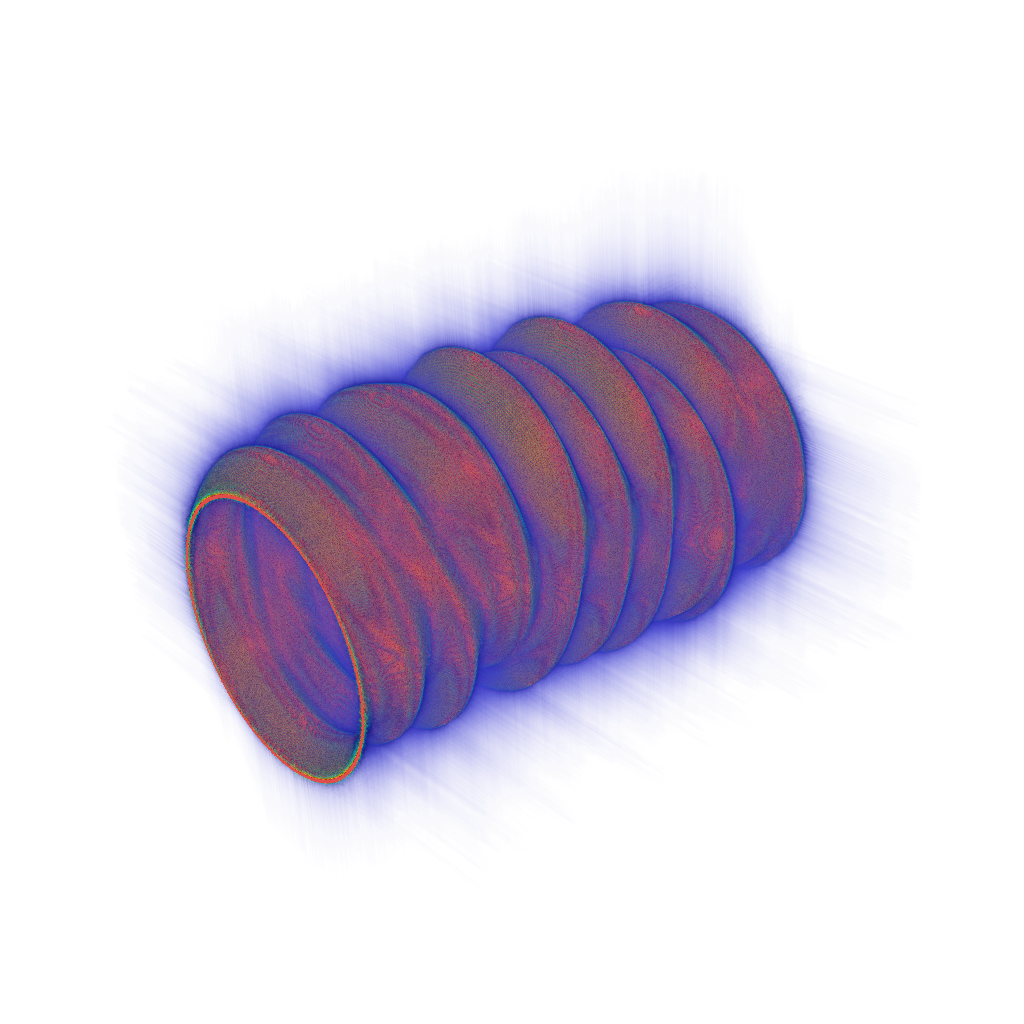}
    \end{subfigure}
    \begin{subfigure}{0.45\textwidth}
        \includegraphics[width=\textwidth,trim={100 100 100 100},clip]{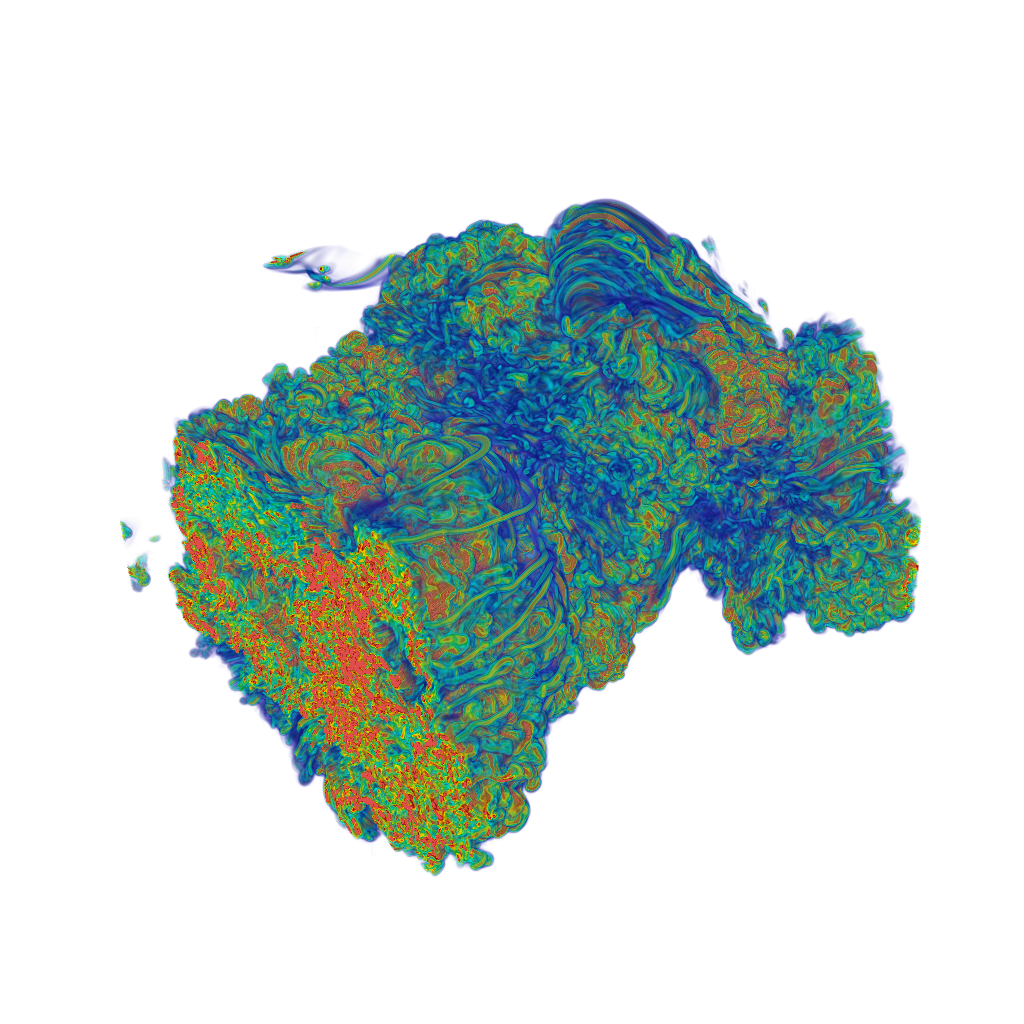}
    \end{subfigure}
    \caption{Vorticity magnitude of the Cylindrical Shear Flow at time $t = 0$ (left) and time $t = 1$ (right) at a resolution of $N = 512$. This experiment provides an especially good view of the vortex stretching in turbulent flows, as vortex tubes are clearly visible on the boundary between turbulent and laminar flow regimes.}
    \label{fig:csf_sample}
\end{figure*}

\begin{figure*}
    \foreach \N in {64,128,256,512} {
        \begin{subfigure}{0.225\textwidth}
            \includegraphics[width=\textwidth,trim={100 100 100 100},clip]{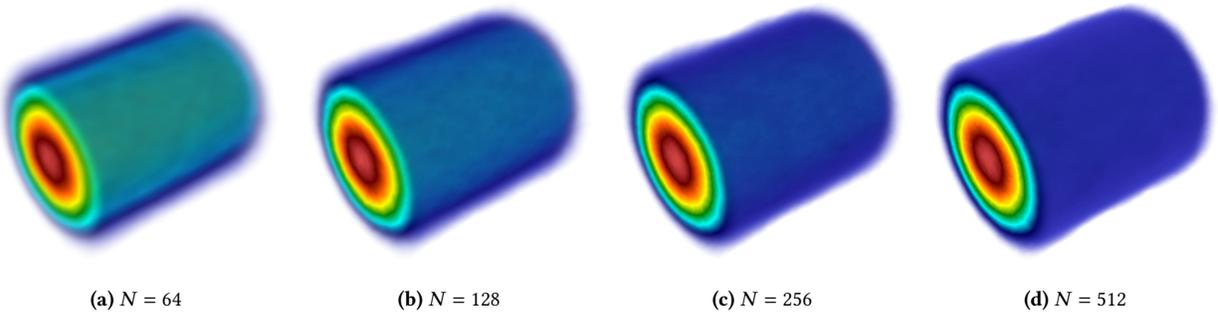}
            \caption{$N = \N$}
        \end{subfigure}
    }
    \caption{Mean $x$-component of the flow field for the Cylindrical Shear Flow at time $t = 1$ at different resolutions.}
    \label{fig:csf_mean}
\end{figure*}

\begin{figure*}
    \foreach \N in {64,128,256,512} {
        \begin{subfigure}{0.225\textwidth}
            \includegraphics[width=\textwidth,trim={100 100 100 100},clip]{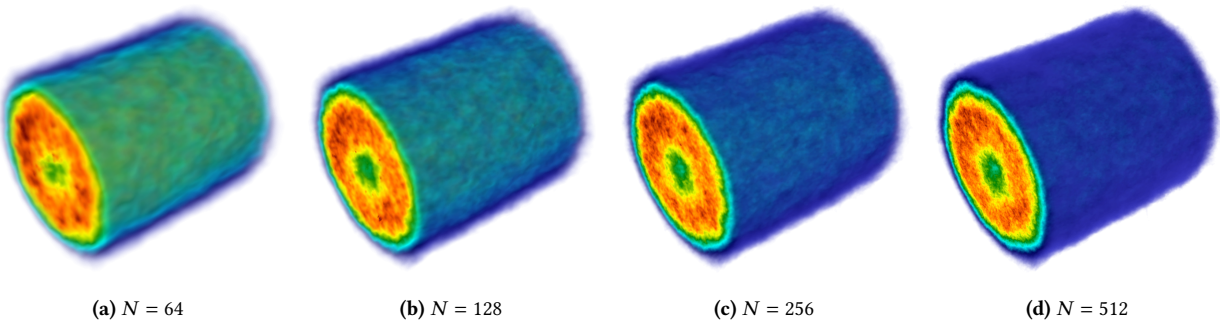}
            \caption{$N = \N$}
        \end{subfigure}
    }
    \caption{Variance of $x$-component of the flow field for the Cylindrical Shear Flow at time $t = 1$ at different resolutions.}
    \label{fig:csf_var}
\end{figure*}

The simulations again use hyperviscosity parameter $s = 1.5$ in order to extend the intermittent range. A single realization of the initial and final conditions for $\rho = 0$ is shown in Figure \ref{fig:csf_sample} where we plot the vorticity magnitude of the flow field. As expected, turbulence spreads from the initial shear layer outward giving a beautiful view into its delicate structures. Likewise, we visualize the mean (Figure \ref{fig:csf_mean}) and the variance (Figure \ref{fig:csf_var}) for different mesh resolutions. These low moments of the statistical solution contain only relatively low frequencies resulting in their smeared out appearance. This also suggests that in theory only few Fourier modes are needed to accurately approximate the mean and variance of statistical solutions and hints at the possibility of quite accurate simulations thereof with e.g. sophisticated turbulence modeling.

\section{Conclusions}

We have successfully implemented and optimized a spectral hyperviscosity solver for the incompressible Navier-Stokes equations and their vanishing viscosity limit. The solver's efficiency enables computation of both two- and three-dimensional large-scale statistical solutions to fluid flows which in turn can be used to deepen our knowledge about turbulence and serve as an invaluable source of training data for scientific Machine Learning models. Benchmarks of the code show close to optimal performance on the Piz Daint supercomputer, while also having excellent strong and weak scaling properties.

Optimality of the computational kernels was assessed by comparing their runtimes against their analytically computed lower bounds. These bounds assume zero cost for arithmetic operations and use memory bandwidths measured on the Piz Daint cluster. Furthermore, all additional simplifications were chosen such that they introduce a bias toward lower runtimes. Nonetheless, the MPI transpose responsible for half of the total runtime of the simulation still gets over 50\% efficiency in these highly conservative estimates. Even more notably, it beats the speed of the native MPI implementation on Piz Daint by almost 20\%.

We present the results of two simulations performed using our code. The well-known Taylor-Green vortex \cite{TaylorGreen1937} demonstrates the need for statistical solutions excellently by developing turbulence throughout the domain early in the simulation and consequently not showing any sample wise convergence under mesh refinement. Statistical quantities on the other hand converge with a good rate making them useful in further research on turbulence. Additionally, the experiment clearly demonstrates Kolmogorov's hypothesized energy dissipation in the vanishing viscosity limit of the Navier-Stokes equations.
As a second experiment, the Cylindrical Shear Flow demonstrates the statistical solutions ability to capture properties of flow fields with highly varying turbulence regimes accurately. Particular focus is put on the demonstration of the approximation capabilities on the pointwise mean and variance as they belong to the most important quantities used to conceptualize probability distributions.

The work introduced here paves the way for considerably more research into the behavior of turbulent fluid flows, especially their statistical properties. Combined with the convergence of statistical solutions under mesh refinement, this provides a promising path towards more accurate solutions to turbulent fluids than possible with current turbulence modeling strategies.

\begin{acks}
This work was supported by a grant from the Swiss National Supercomputing Centre (CSCS) under project ID 1217.
\end{acks}

\printbibliography

%\appendix

\end{document}